\documentclass[eqsecnum,aps,epsfig]{revtex4}
\usepackage{amsmath,amssymb,amsthm,bm}
\usepackage{psfrag}
\usepackage[dvips]{graphicx}
\usepackage[utf8]{inputenc} 
\usepackage{hyperref}
\usepackage{xcolor}  
\usepackage{array}   
\usepackage{amsmath} 
\usepackage{hyperref}
\usepackage{booktabs}

\begin{document}
\title{Domain Structures and Static Potentials in \(G_2\) Gauge Theory}

\author{Seyed Mohsen Hosseini Nejad}
\email{smhosseininejad@semnan.ac.ir}
\affiliation{
Faculty of Physics, Semnan University, P.O. Box 35131-19111, Semnan, Iran}

\begin{abstract}

We investigate the domain structures of the \(G_2\) vacuum and the associated static potentials within the domain model, considering both zero and non-zero aggregate flux vacuum domains. Potentials associated with domains of zero aggregate flux exhibit favorable Casimir scaling and convexity properties at intermediate distances, and their long-range behavior matches the expected ordering of constant potential values according to representation. In contrast, potentials derived from domains with nonzero aggregate flux deviate from the key features of the confining force, as the absence of a minimal center-flux magnitude renders such configurations physically improbable. Furthermore, the potential ratios for \(SU(3)\) and \(SU(2)\) subgroups are found to be consistent with Casimir ratios of \(G_2\) at intermediate regimes, suggesting a possible role for these subgroups in the confinement mechanism of the \(G_2\) gauge group.
 \\  \\     
\end{abstract}

\maketitle

\section{INTRODUCTION}\label{Sect0}

Understanding the mechanism of color confinement remains a fundamental challenge in quantum chromodynamics (QCD). Numerical simulations of SU($N$) lattice gauge theories with non-trivial center elements support the Casimir scaling hypothesis, which posits that the static potential between color sources at intermediate distances is proportional to the eigenvalues of the quadratic Casimir operator \cite{Deldar:1999vi,Bali:2000un,Piccioni:2005un}. At asymptotic distances, the potential is governed by a string tension determined exclusively by $N$-ality, matching the value for the lowest representation in the same class \cite{Kratochvila:2003zj}. Furthermore, the static potential is constrained to be a convex function at all distances, precluding any concave regions \cite{Bachas:1985xs}.

Lattice simulations and infrared models indicate that center vortices \cite{tHooft:1977nqb,Vinciarelli:1978kp,Yoneya:1978dt,Cornwall:1979hz,Mack:1978rq,Nielsen:1979xu,Oxman:2010jk,Dehghan:2024rly,Asmaee:2021xkm,Junior:2019fty,Golubich:2021kjc,Schweigler:2012ae,Hayashi:2025doq,Hollwieser:2012kb,Nejad:2024,Dehghan:2024blw}, quantized magnetic flux tubes defined by non-trivial center elements, play a key role in confinement. These vortices induce an area-law behavior in the Wilson loop, resulting in a linearly rising potential in the infrared regime. Furthermore, numerical simulations indicate that center vortices also explain key phenomena related to chiral symmetry, including the generation of topological charge and the mechanism of spontaneous chiral symmetry breaking \cite{ schweigler:2012ae,engelhardt:2002qs,hollwieser:2012kb,HosseiniNejad:2015oeu,hollwieser:2011,HosseiniNejad:2016fcl,leinweber:2006zq,hollwieser:2008tq,Faber:2017alm}.

The domain structure model provides a framework for explaining color confinement through the interaction of Wilson loops with center domains, focusing on a two-dimensional plane in (Euclidean) spacetime \cite{Faber:1997rp,Deldar:2001,Greensite:2006sm,Deldar:2007,Nejad:2014tja,Nejad:2019}. In this model, the QCD vacuum is assumed to consist of two types of domains: those associated with non-trivial center elements (center vortex type) and those corresponding to the identity element (vacuum type).

Despite the successes of the center vortex model, the relationship between confinement and center symmetry remains incompletely understood. To clarify the role of center elements, a promising approach involves studying confinement in gauge groups that lack nontrivial center elements, such as \(G_2\). The exceptional Lie group \(G_2\), the smallest of the exceptional simple Lie groups, naturally arises as a subgroup of SO($7$) and serves as an ideal candidate for such investigations. Lattice calculations in \(G_2\) gauge theory \cite{Liptak:2008gx,Wellegehausen:2011} reveal a linear potential for various representations at intermediate distances, consistent with Casimir scaling. While this behavior is well-established numerically, the theoretical basis for the emergence of a linear potential at intermediate distances remains unclear. Unlike SU($N$) gauge theories, which typically confine fundamental charges, \(G_2\) exhibits qualitatively different behavior: all of its representations, including the fundamental representation, eventually become screened through gluon dynamics at sufficiently large distances. In our previous work, \cite{Deldar2012}, we investigated the static potentials in \(G_2\) gauge theory using a simplified vacuum domain model. That model considered only one type of vacuum domain, namely those with non-zero aggregate flux. The angle function was described by a simple ansatz, and only a single Cartan generator of \(G_2\) was used in the construction of the domain cores. As a result, this approach led to unphysical concavities in the static potentials and showed only limited agreement with Casimir scaling predictions.

In the present work, we generalize the vacuum domain model by adopting a square ansatz for the angle function, which allows us to study both zero and non-zero aggregate flux domains. In addition, we employ both Cartan generators of \(G_2\) in the domain construction. These extensions provide a more comprehensive description of the static potential, allowing us to characterize its properties in greater detail. Potentials associated with domains of zero aggregate flux exhibit favorable Casimir scaling and convexity properties at intermediate distances, and their long-range behavior reproduces the expected ordering of the asymptotic potential values with respect to the representation, while those of non-zero aggregate flux deviate from this behavior.

Although \(G_2\) has a trivial center and thus no conventional center vortices, it nevertheless exhibits a confining regime. Lattice simulations \cite{Greensite:2006sm} indicate that removing gauge links associated with the \(SU(3)\) and \(SU(2)\) subgroups of \(G_2\) either significantly alters the full \(G_2\) confinement behavior or completely eliminates confinement signatures. Furthermore, in \(G_2\) gauge-Higgs theory \cite{Holland2003}, introducing a Higgs field in the $\{7\}$ representation induces spontaneous symmetry breaking \(G_2\) to \(SU(3)\), giving mass to 6 of the 14 gluons while preserving the masslessness of the remaining 8 gluons associated with the \(SU(3)\) subgroup. Our analysis suggests a possible role for center vortices originating from the \(SU(3)\) and \(SU(2)\) subgroups of \(G_2\) in the intermediate confinement regime. To investigate this, we fix the model parameters by fitting the $\{7\}$ representation potentials of \(G_2\), \(SU(3)\), and \(SU(2)\) to be nearly parallel at intermediate distances. With these fitted parameters, the potentials for the higher representations ($\{14\}$ and $\{27\}$) of each subgroup are found to remain nearly parallel to those of \(G_2\), consistent with the Casimir scaling behavior observed in \(G_2\) at intermediate distances.

In Section \ref{Sect1}, we implement the domain structure model for the exceptional group \(G_2\). Building on this framework, Section \ref{Sect2} presents our analysis of static potentials in the lowest representations and their ratios, generated by two distinct types of vacuum domains in \(G_2\) gauge theory. Section \ref{Sect3} examines the possible roles of the \(SU(3)\) and \(SU(2)\) subgroups of \(G_2\) in the intermediate regime of these static potentials. Finally, Section \ref{Sect4} provides a comprehensive summary of our key findings. 

\section{The domain model in \(G_2\) gauge group}\label{Sect1}

The domain model of confinement provides a framework to explain key features of the confining force, as supported by lattice simulations \cite{Faber:1997rp,Greensite:2006sm}. In this model, the SU($N$) domains are classified into two types: center vortex domains, associated with non-trivial center elements  $z_n=\exp(i2\pi n/N)$ for $n=1,...,N-1$, and vacuum domains, associated with the trivial center element $z_n=1$ ($n=0,N$). Now, we apply the model to the \(G_2\) gauge group. The homotopy group
\begin{equation}
\Pi_1[G_2/\mathbb{I}] = \mathbb{I},
\end{equation}
implies that \(G_2\) gauge theory has only vacuum domains corresponding to the trivial center element. When contained entirely within the Wilson loop, these vacuum domains do not contribute a non-trivial phase. Nevertheless, there is more than one way to realize the aforementioned phase property: there may be vacuum domains with zero net magnetic flux, called $z_0$ domains, or vacuum domains with non-zero net magnetic flux whose magnitude, when exponentiated as $\exp(i2\pi)$, again yields a unit phase in the evaluation of the Wilson loop. These are denoted $z_1$ domains.

The embedding of \(G_2\) in the group SO($7$) is generated by the 14 Lie algebra elements $H_k$, $k = 1, \dots, 14$. The \(G_2\) group can be fully covered by six distinct \(SU(2)\) subgroups \cite{Cossum2007}. The first three subgroups,
\begin{equation}
\begin{aligned}
&1.\quad H_1,\ H_2,\ H_3, \\
&2.\quad H_4,\ H_5,\ \frac{1}{2} \left( \sqrt{3} H_8 + H_3 \right), \\
&3.\quad H_6,\ H_7,\ \frac{1}{2} \left( - \sqrt{3} H_8 + H_3 \right),
\label{H_123}
\end{aligned}
\end{equation}
form non-reducible, four-dimensional real representations. Collectively, they generate an \(SU(3)\) subgroup of \(G_2\), which is seven-dimensional and reducible. The remaining three \(SU(2)\) subgroups,
\begin{equation}
\begin{aligned}
&4.\quad \sqrt{3} H_8,\ \sqrt{3} H_{9},\ \sqrt{3} H_{10}, \\
&5.\quad \sqrt{3} H_{11},\ \sqrt{3} H_{12},\ \frac{1}{2} \left( - \sqrt{3} H_8 + 3 H_3 \right), \\
&6.\quad \sqrt{3} H_{13},\ \sqrt{3} H_{14},\ \frac{1}{2} \left( \sqrt{3} H_8 + 3 H_3 \right),
\label{H_456}
\end{aligned}
\end{equation}
act via seven-dimensional reducible representations. The generators $H_3$ and $H_8$ are the Cartan generators of the full \(G_2\) algebra.

The effect of a vacuum domain of type $n$ on a planar Wilson loop in representation $r$ is to multiply the loop by a group factor:
\begin{equation}
{G}_{\{r\}}(\alpha^{n})=\frac{1}{d_r}Tr\left(\exp\left[i({\alpha}_3^{n}{H}_3^{\{r\}}+{\alpha}_8^{n}{H}_8^{\{r\}})\right]\right),
\label{group_factor}
\end{equation}
where $d_r$ is the dimension of representation $r$, ${H}_3^{\{r\}}$ and ${H}_8^{\{r\}}$ are Cartan generators of \(G_2\), and the angles ${\alpha}_i^{n}$ ($i=3,8$) depend on both the Wilson loop and the domain's position $x$. The Cartan generators for the lowest representations of \(G_2\) are provided in the Appendix.

The influence of a vacuum domain on a Wilson loop depends on its position relative to the loop's area. If a vacuum domain is entirely contained within the Wilson loop, the condition
\begin{equation}
\exp\left[i({\alpha}_3^{n}{H}_3^{\{r\}}+{\alpha}_8^{n}{H}_8^{\{r\}})\right]=\mathbb{I}_{d_r},
\label{max_alpha}
\end{equation}
is satisfied, and the vacuum domain has no effect on the loop. This constraint can be used to determine the maximum value of the angle, $\alpha^{n}_{\text{max}}$. Conversely, if the Wilson loop encloses only a fraction of a vacuum domain, the loop is affected.

The static quark potential induced by the domains is given by \cite{Faber:1997rp,Greensite:2006sm}:
\begin{equation}
\label{potential}
V_{\{r\}}(R) = -\sum_{x}\ln( 1 - \sum_{n} f_{n}
[1 - {\mathrm {Re}}{G}_{\{r\}} ({\alpha}^{n}(x))]),
\end{equation}
where $f_n$ denotes the probability that a given plaquette is pierced by a domain of type $n$.

\begin{figure}[h!]
\centering
\includegraphics[width=0.28\columnwidth]{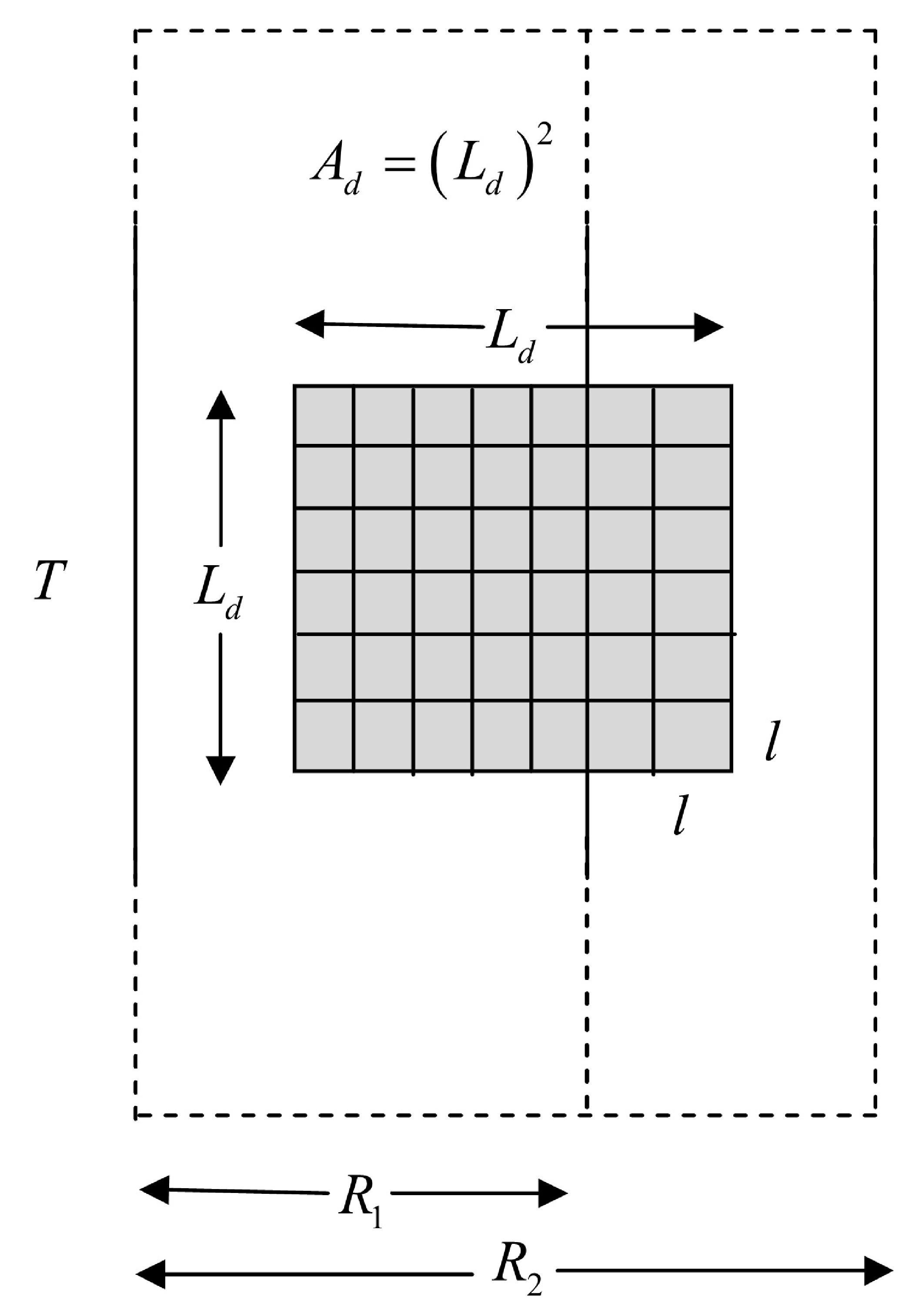}
\caption{Schematic illustration of the effect of a vacuum domain core, modeled via the square ansatz, on Wilson loops of different extents. The $R_1 \times T$ loop encloses a fraction of the vacuum domain, acquiring a non-trivial group factor $\mathrm{Tr}[U \cdots (G \mathbb{I}) \cdots U]$. In contrast, the $R_2 \times T$ loop fully contains the vacuum domain, which consequently contributes trivially as $\mathrm{Tr}[U \cdots \mathbb{I} \cdots U]$ and leaves the loop unchanged.}\label{fig:0}
\end{figure}

The square ansatz for the angle $\alpha^{n}(x)$ was introduced by Greensite \textit{et al.} \cite{Greensite:2006sm}. In this model, a domain of cross-sectional area $A_d$ is partitioned into subregions of area $l^2 \ll A_d$, where $l$ is a short correlation length. The color magnetic fluxes within these subregions fluctuate randomly and nearly independently, resulting in uncorrelated fluxes between neighboring subregions. A key constraint is that the net color magnetic flux from all subregions must correspond to an element of the gauge group center. Within this framework, the square ansatz for the angle is given by:
\begin{equation}
\label{Sansax}
          \left({\alpha}_i^{n}(x)\right)^2 = \frac{A_d}{ 2\mu} \left[
\frac{A}{ A_d} - \frac{A^2}{ A_d^2} \right]
                         + \left(\alpha^{n}_{max} \frac{A}{ A_d}\right)^2,
\end{equation}
where $A$ is the cross-sectional area of the center domain overlapping with the minimal area of the Wilson loop, $\mu$ is a free parameter, and the domain is characterized by a square cross-section of area $A_d = L_d^2$. The square ansatz requires evaluating the function ${\alpha}_i^{n}(x)$ over two distinct intervals of $R$, defined relative to the domain size $L_d$. A schematic illustration of the interaction between the vacuum domain with square ansatz angle and the Wilson loops is presented in Fig. \ref{fig:0}.
The parameter $x$ is restricted to the interval $-\frac{L_d}{2} \le x \le R + \frac{L_d}{2}$. This range encompasses all plaquettes within the minimal area of the Wilson loop, as well as those plaquettes outside the loop's perimeter but within a distance $L_d/2$ of it. The expressions for the square ansatz are then given by:

For $0 \le R \le L_d$:
    \begin{equation}
    ({\alpha}_i^{n}(x))^2 = 
    \begin{cases} 
    \dfrac{L_d^2}{2\mu} \left( \dfrac{y}{L_d} - \dfrac{y^2}{L_d^2} \right) + \left( \alpha^n_{\text{max}} \dfrac{y}{L_d} \right)^2, & \text{for } -\dfrac{L_d}{2} \le x \le -\dfrac{L_d}{2} + R, \\[10pt]
    \dfrac{L_d^2}{2\mu} \left( \dfrac{R}{L_d} - \dfrac{R^2}{L_d^2} \right) + \left( \alpha^n_{\text{max}} \dfrac{R}{L_d} \right)^2, & \text{for } -\dfrac{L_d}{2} + R \le x \le \dfrac{L_d}{2}, \\[10pt]
    \dfrac{L_d^2}{2\mu} \left( \dfrac{y}{L_d} - \dfrac{y^2}{L_d^2} \right) + \left( \alpha^n_{\text{max}} \dfrac{y}{L_d} \right)^2, & \text{for } \dfrac{L_d}{2} \le x \le R + \dfrac{L_d}{2}.
    \end{cases}
    \label{eq:ansatz_short}
    \end{equation}

For $L_d \le R$:
    \begin{equation}
    ({\alpha}_i^{n}(x))^2 = 
    \begin{cases} 
    \dfrac{L_d^2}{2\mu} \left( \dfrac{y}{L_d} - \dfrac{y^2}{L_d^2} \right) + \left( \alpha^n_{\text{max}} \dfrac{y}{L_d} \right)^2, & \text{for } -\dfrac{L_d}{2} \le x \le \dfrac{L_d}{2}, \\[10pt]
    \left( \alpha^n_{\text{max}} \right)^2, & \text{for } \dfrac{L_d}{2} \le x \le R - \dfrac{L_d}{2}, \\[10pt]
    \dfrac{L_d^2}{2\mu} \left( \dfrac{y}{L_d} - \dfrac{y^2}{L_d^2} \right) + \left( \alpha^n_{\text{max}} \dfrac{y}{L_d} \right)^2, & \text{for } R - \dfrac{L_d}{2} \le x \le R + \dfrac{L_d}{2}.
    \end{cases}
    \label{eq:ansatz_long}
    \end{equation}

The spatial dependence in the above expressions is governed by the function $y(x)$, which determines the effective overlap width:
\begin{equation}
y(x) = 
\begin{cases} 
R - x + \dfrac{L_d}{2}, & \text{for } |R - x| \le |x|, \\[10pt]  %
x + \dfrac{L_d}{2}, & \text{for } |R - x| > |x|.
\end{cases}
\label{eq:y_definition}
\end{equation}

\begin{figure}[h!]
\centering
a)\includegraphics[width=0.43\columnwidth]{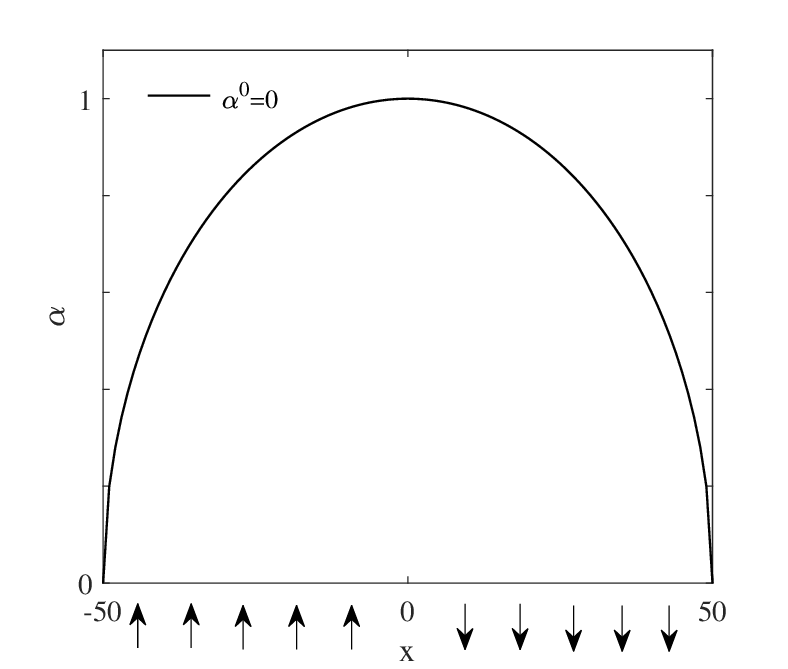}
\hspace{3em}
b)\includegraphics[width=0.43\columnwidth]{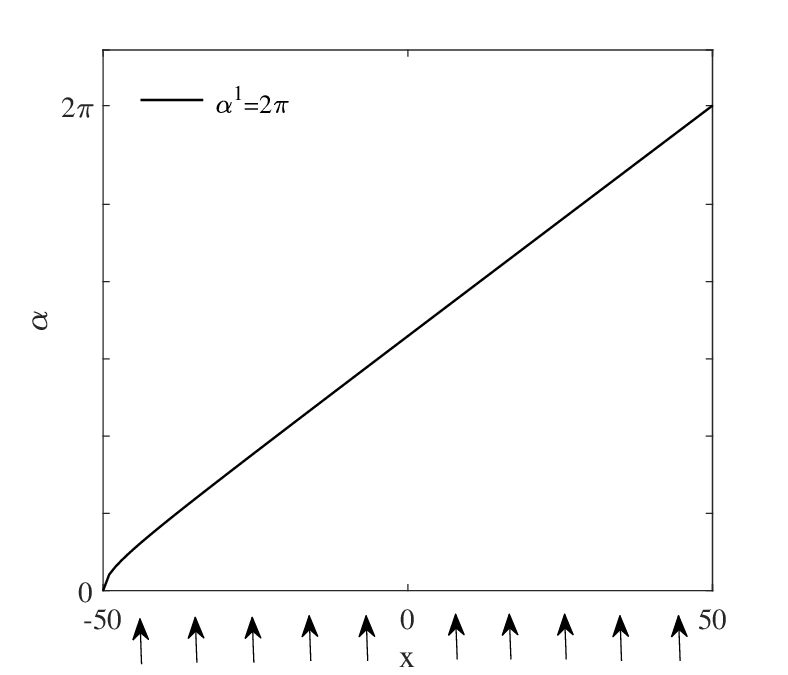}
\caption{Angle \(\alpha(x)\) for a Wilson loop with left and right legs positioned at \(x = 0\) and \(x = R = 100\), shown near the left leg (at \(x = 0\)) for the vacuum domains: a) with zero net magnetic flux (\(z_0\) vacuum domains); b) with non-zero net magnetic flux (\(z_1\) vacuum domains). The up and down arrows indicate the direction of the partial flux of the vacuum domain, which is gradually added to the minimal area of the Wilson loop as \(x\) increases. The free parameters are \(L_d = 100\) and \(L_d^2/(2\mu) = 4\).}\label{fig:alpha}
\end{figure}

Figures \ref{fig:alpha}(a) and (b) show the square ansatz for the angle $\alpha(x)$ of $z_0$ and $z_1$ vacuum domains, respectively, with the left and right legs of the Wilson loop positioned at $x = 0$ and $x = R = 100$.
The plot shows the angles near the left leg of the Wilson loop (at $x = 0$). This range, $-L_d/2 \le x \le L_d/2$, encompasses the plaquettes around $x = 0$ within a distance $L_d/2$. The free parameters are $L_d = R= 100$ and $L_d^2/(2\mu) = 4$. The arrows show the direction of a partial flux of the vacuum domain, which is gradually added to the minimal area of the Wilson loop as $x$ increases. For the angle $\alpha(x)$ of $z_1$ vacuum domains, from $-50$ to $50$, the flux of the domain located in the Wilson loop increases gradually from 0 to $2\pi$. The up arrows show that the flux in a $z_1$ vacuum domain has the same direction in all thickness, $L_d = 100$. For the angle $\alpha(x)$ of $z_0$ vacuum domains, from $-50$ to $0$, the flux of the domain located in the Wilson loop increases gradually (up arrows) and then from $0$ to $50$, it decreases (down arrows), resulting in zero net magnetic flux. In other words, half of the flux of a $z_0$ vacuum domain points in one direction, while the other half points in the opposite direction. 

In the next section, we focus on the contributions of both types of vacuum domains to the static potentials.

\section{Casimir scaling and color screening}\label{Sect2}

For the \(G_2\) gauge group, which possesses only a trivial center, there may exist vacuum domains with non-zero net magnetic flux ($z_1$ domains) or vacuum domains with zero net magnetic flux ($z_0$ domains).

First, we study a model featuring only $z_1$ domains. Using Eq. (\ref{potential}), the static potential induced by $z_1$ vacuum domains in the \(G_2\) gauge group is given by:
\begin{equation}
V_{\{r\}}(R) =- \sum^{{ {L_d}/2 + R}}_{{x=-{L_d}/2}} \ln[(1-f_1) + f_1{\mathrm {Re}}{G}_{\{r\}}(\alpha^{1}(x))], 
\end{equation}
where $f_1$ is the probability that any given plaquette is pierced by a $z_1$ vacuum domain. For any irreducible representation $r$ of a Lie algebra, the trace of the product of two generators satisfies:
\begin{equation}
\mathrm{Tr}(H^a H^b) = \frac{C_2(r) \dim(r)}{\dim(\text{adj})} \delta^{ab},
\end{equation}
where $H^a$ denotes a generator in representation $r$, $C_2(r)$ is the quadratic Casimir operator value, and $\dim(r)$ and $\dim(\text{adj})$ represent the dimensions of representation $r$ and the adjoint representation, respectively.

\begin{table}[htbp]
\centering
\caption{Properties of low-dimensional representations of \(G_2\)}
\label{tab:1}
\begin{tabular}{l@{\hspace{2em}}c@{\hspace{2em}}c@{\hspace{2em}}c}  
\hline
Property & \{7\} (Fundamental) & \{14\} (Adjoint) & \{27\} \\
\hline
Quadratic Casimir ($C_2$) & 2 & 4 & $\frac{14}{3}$ \\[0.5ex]
Casimir Ratio & 1 (reference) & 2 & $\frac{7}{3} \approx 2.33$ \\[0.5ex]
Trace Normalization & $\mathrm{Tr}(H_aH_b) = \frac{1}{2}\delta_{ab}$ & $\mathrm{Tr}(H_aH_b) = 2\delta_{ab}$ & $\mathrm{Tr}(H_aH_b) = \frac{9}{2}\delta_{ab}$ \\[1.5ex]

\hline
\end{tabular}
\end{table}

Table \ref{tab:1} summarizes some properties of the lowest-dimensional \(G_2\) representations. Using the decompositions of these \(G_2\) representations into \(SU(3)\) subgroup representations, one can construct the Cartan generators of \(G_2\) from those of \(SU(3)\) subgroup; these are provided in the Appendix. Using the Cartan generators for the \(G_2\), we calculate the real part of the group factor defined in Eq. (\ref{group_factor}); the results are also presented in the Appendix.

The square ansatz given in Eq. (\ref {Sansax}) for the angles corresponding to the $z_1$ vacuum domains, applicable to all representations, takes the form:
\begin{equation}
         ({\alpha}_3^{1}(x))^2 = \frac{A_d}{ 2\mu} \left[
\frac{A}{ A_d} - \frac{A^2}{ A_d^2} \right],~~~
 \\
         ({\alpha}_8^{1}(x))^2 = \frac{A_d}{ 2\mu} \left[
\frac{A}{ A_d} - \frac{A^2}{ A_d^2} \right]
                         + \left(2\pi\sqrt{24}\frac{A}{ A_d}\right)^2.    
\label{ansax_z_1}                                               
\end{equation}

\begin{figure}[h!]
\centering
a)\includegraphics[width=0.43\columnwidth]{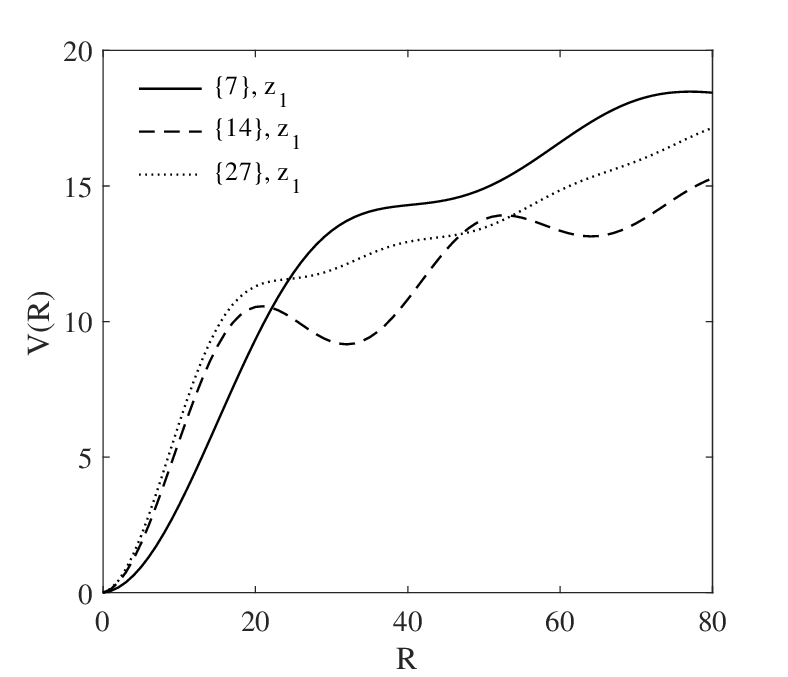}
\hspace{3em}
b)\includegraphics[width=0.43\columnwidth]{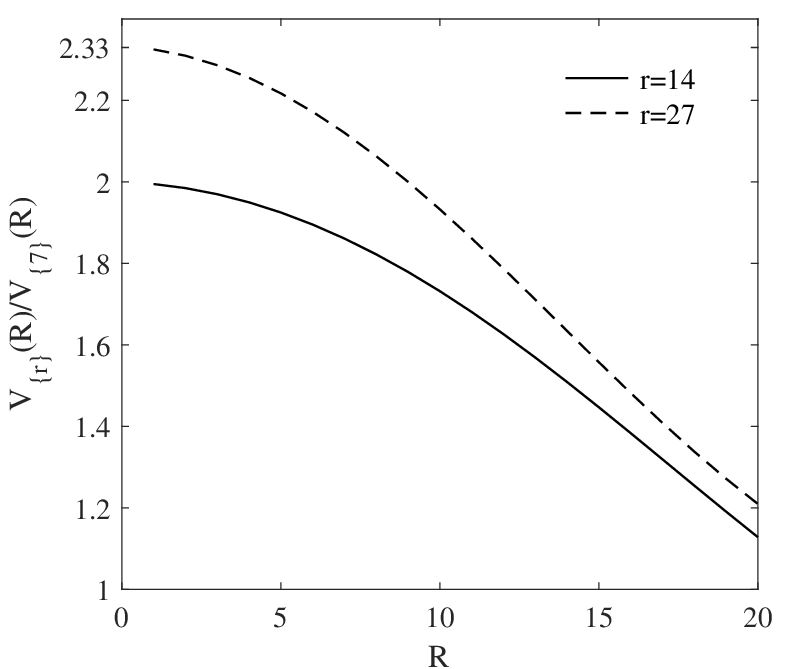}
\caption{a) Static potentials induced by $z_1$ vacuum domains. The potentials exhibit linear behavior for $R \in [0,20]$, where the vacuum domain is partially enclosed within the Wilson loop, but display unphysical concavities in the intermediate regime that influence the connection to long-distance behavior. b) Ratios $V_{\{r\}}(R)/V_{\{7\}}(R)$ induced by $z_1$ vacuum domains at intermediate distances. These ratios change rapidly relative to the expected Casimir ratios and deviate significantly from exact Casimir scaling. The parameters are $L_d = 100$, $f_1 = 0.1$, and $L_d^2/(2\mu) = 4$.}\label{fig:1}
\end{figure}

We set the free parameters to $L_d = 100$, $f_1 = 0.1$, and $L_d^2/(2\mu) = 4$, with a correlation length $l = 1$. This choice causes the linear potentials to originate from $R = l$. Figure \ref{fig:1}(a) shows the static potentials $V_{\{r\}}(R)$ induced by $z_1$ vacuum domains for the fundamental (${\{7\}}$), adjoint (${\{14\}}$), and ${\{27\}}$ representations, computed over the range $R \in [0,80]$. 

At intermediate distances $R \in [0,20]$, the potentials induced by $z_1$ vacuum domains exhibit linear behavior. However, the computed static potentials display unphysical concavity within this intermediate regime, which connects to the long-distance behavior. Figure \ref{fig:1}(b) shows the potential ratios $V_{\{14\}}(R)/V_{\{7\}}(R)$ and $V_{\{27\}}(R)/V_{\{7\}}(R)$ induced by $z_1$ vacuum domains. These ratios initially coincide with the Casimir ratios given in Table \ref{tab:1}, but decrease rapidly over the range $R \in [1,20]$.

In Ref. \cite{Deldar2012}, the potentials for $z_1$ vacuum domains were computed using an alternative profile function for the angle $\alpha$ and considering only the contribution of the Cartan generator $H_8$. That approach yielded potentials that deviated from exact Casimir scaling and exhibited unphysical concavities.

\begin{figure}[h!]
\centering
\includegraphics[width=0.43\columnwidth]{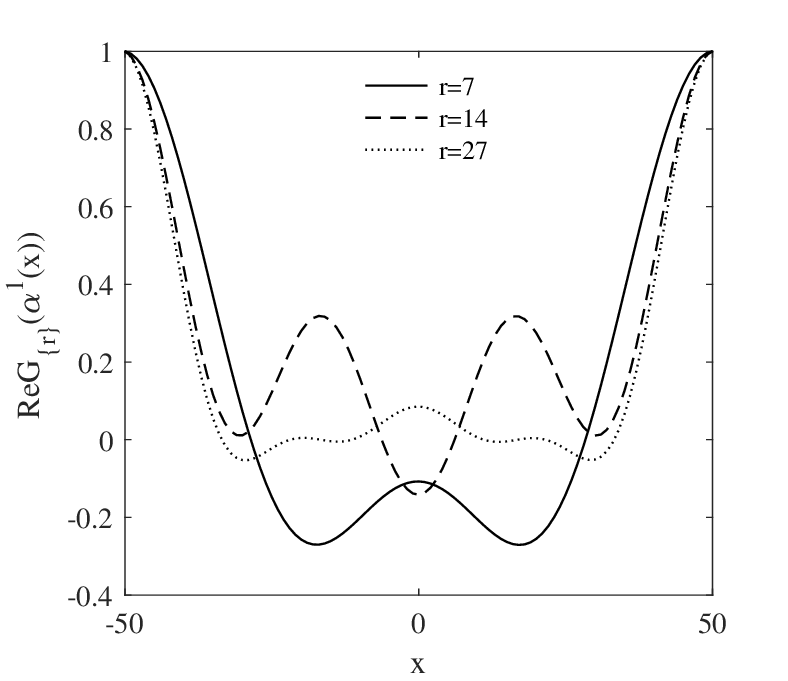}
\caption{ Real parts of the group factors $\mathrm{Re} G_{\{r\}}(\alpha)$ for $z_1$ vacuum domains, plotted as functions of $x$. The plot shows oscillatory behavior near the left time-like leg of the Wilson loop (at $x = 0$); an identical pattern appears near the right leg (at $x = R=100$). This oscillatory structure at the boundaries is consistent across all representations shown. The free parameters are $L_{d}=100$ and $L^{2}_{d}/(2\mu)=4$.}\label{fig:1-1}
\end{figure}

Now, using the square ansatz and incorporating both Cartan generators $H_3$ and $H_8$, we find that the concavities in the potentials induced by  $z_1$ vacuum domains persist. In detail, Fig. \ref{fig:1-1} shows the real part of the group factor, $\mathrm{Re} G_r(\alpha^n(x))$, induced by $z_1$ vacuum domains for the specified representations, with the left and right legs of the Wilson loop positioned at $x = 0$ and $x = R=100$, respectively. As shown, the group factors exhibit oscillatory behavior near the left leg of the Wilson loop (at $x = 0$); an equivalent oscillatory pattern is also observed near the right leg (at $x = 100$) for the representations. This indicates that the concavities observed in the potentials originate from the oscillatory characteristics of their corresponding group factors.

Consequently, a model featuring only $z_1$ domains violates the convexity of the static potential and does not display favorable Casimir scaling properties. 

It is essential to obtain convex potentials, free from concavities, to ensure a physically consistent interpolation between intermediate and asymptotic regimes. Furthermore, the potentials should exhibit a gradual deviation from exact Casimir scaling at intermediate distances.

\begin{figure}[h!]
\centering
a)\includegraphics[width=0.43\columnwidth]{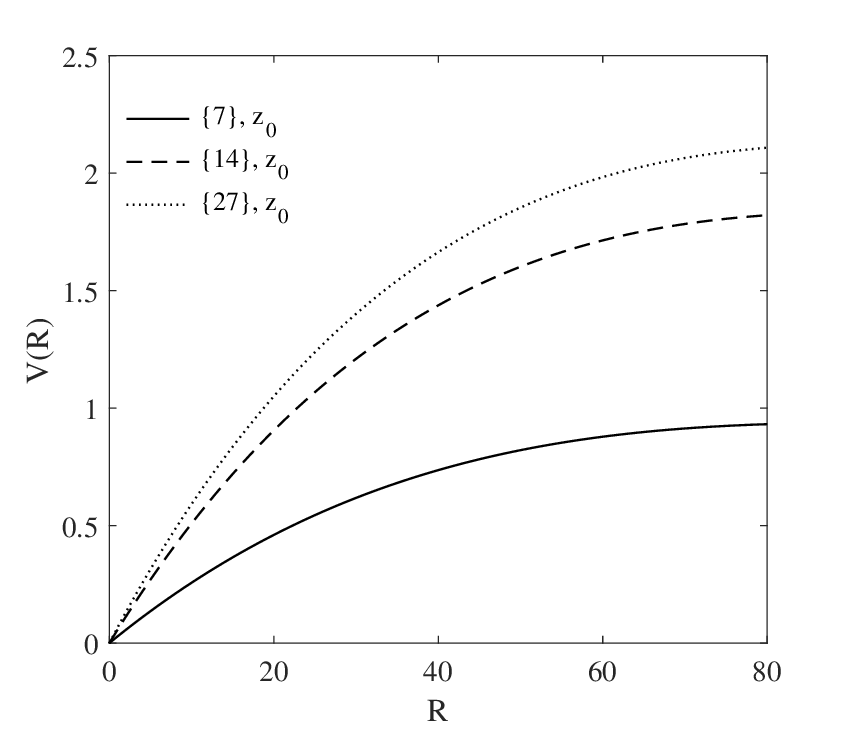}
\hspace{3em}
b)\includegraphics[width=0.43\columnwidth]{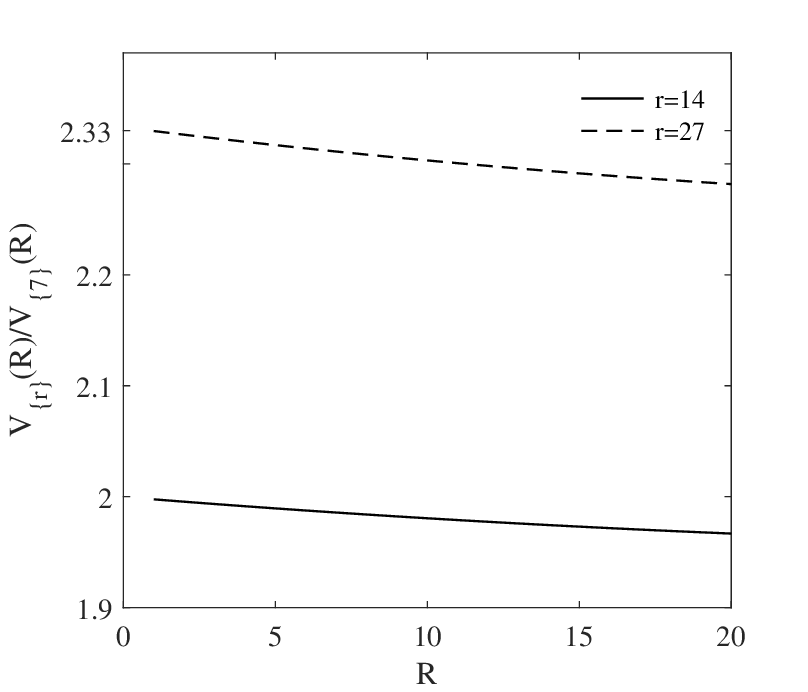}
\caption{a) Static potentials induced by $z_0$ vacuum domains. The potentials are linear in the range $R\in [0,20]$, where the $z_0$ vacuum domain is partially enclosed within the Wilson loop. b) Potential ratios 
$V_{\{r\}}(R)/V_{\{7\}}(R)$ induced by $z_0$ vacuum domains at intermediate distances. These potentials agree with nearly exact Casimir scaling when the vortex flux located in the medium-sized loops is typically small. The parameters are $L_{d}=100$, $f_{0}=0.1$, and $L^{2}_{d}/(2\mu)=4$.}\label{fig:2}
\end{figure}

Now, we study a model featuring only $z_0$ domains. Using Eq. (\ref{potential}), the static potential induced by $z_0$ vacuum domains in the \(G_2\) gauge group is given by:
\begin{equation}
V_{\{r\}}(R) =- \sum^{{ {L_d}/2 + R}}_{{x=-{L_d}/2}} \ln[(1-f_0) + f_0{\mathrm {Re}}{G}_{\{r\}}(\alpha^{0}(x))], \label{pot:z_0}
\end{equation}
where $f_0$ denotes the probability that a given plaquette is pierced by a $z_0$ vacuum domain. The square ansatz given in Eq. (\ref {Sansax}) for the angles corresponding to $z_0$ vacuum domains, applicable to all representations, takes the form:
\begin{equation}
         ({\alpha}_3^{0}(x))^2 = \frac{A_d}{ 2\mu} \left[
\frac{A}{ A_d} - \frac{A^2}{ A_d^2} \right],~~~
 \\
         ({\alpha}_8^{0}(x))^2 = \frac{A_d}{ 2\mu} \left[
\frac{A}{ A_d} - \frac{A^2}{ A_d^2} \right]. \label{ansax_z0}                       
\end{equation}
We set the free parameters to $L_d = 100$, $f_0 = 0.1$, and $L_d^2/(2\mu) = 4$, with a correlation length $l = 1$. Figure \ref{fig:2}(a) shows the static potentials $V_{\{r\}}(R)$ induced by $z_0$ vacuum domains, plotted for ${\{7\}}$, ${\{14\}}$, and ${\{27\}}$ representations over the range $R \in [0,80]$.

At intermediate distances ($R \in [0,20]$), where the vacuum domain is partially enclosed within the Wilson loop, the potentials exhibit linear behavior and show no concavity. Figure \ref{fig:2}(b) displays the potential ratios $V_{\{14\}}(R)/V_{\{7\}}(R)$ and $V_{\{27\}}(R)/V_{\{7\}}(R)$ induced by $z_0$ vacuum domains. These ratios initially match the Casimir ratios provided in Table \ref{tab:1} and change very slowly over the range $R \in [1,20]$, respectively.

\begin{figure}[h!]
\centering
\includegraphics[width=0.43\columnwidth]{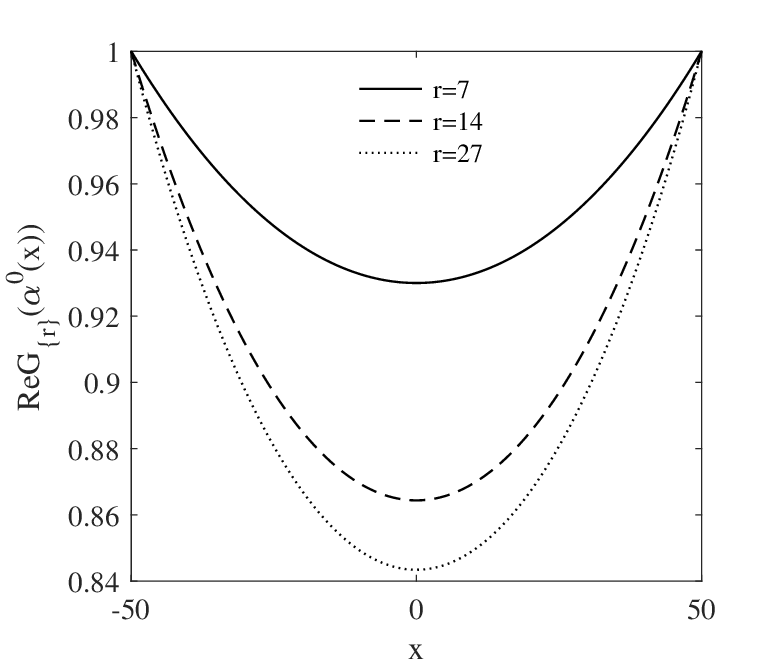}
\caption{Real parts of the group factors $\mathrm{Re}G_r(\alpha)$ for $z_0$ vacuum domains, plotted as functions of $x$. The group factors vary smoothly and remain near unity near the left time-like leg of the Wilson loop (at $x = 0$); identical behavior occurs near the right leg (at $x = R = 100$). This smooth variation near both boundaries is consistently observed across all representations. The free parameters are $L_{d}=100$ and $L^{2}_{d}/(2\mu)=4$.}\label{fig:2-1}
\end{figure}

In detail, Fig. \ref{fig:2-1} displays the real part of the group factor, $\mathrm{Re}G_r(\alpha^n(x))$, induced by $z_0$ vacuum domains for the specified representations. The left and right legs of the Wilson loop are fixed at $x = 0$ and $x = R = 100$, respectively. As shown, the group factors vary gradually and remain close to unity near the left leg of the Wilson loop (at $x = 0$); a similar gradual variation occurs near the right leg (at $x = 100$) across all representations. This demonstrates that the absence of concavities in the potentials results from the smooth variation of the corresponding group factors.

Notably, the potential ratios induced by $z_0$ vacuum domains decrease much more gradually than those induced by $z_1$ vacuum domains. This comparison shows that the $z_0$ domains, which lead to the smallest magnitude of center flux, yield static potentials that adhere strongly to Casimir scaling in the intermediate regime.

In Ref.~\cite{Wellegehausen:2011}, Wellegehausen \emph{et al.} measured the static potential in 3d \(G_2\) gauge theory in a lattice study, where the plateau is determined by the corresponding gluelump mass. For the fundamental representation ${\{7\}}$, a color singlet is formed by combining with three gluons: ${\{7\}} \otimes {\{14\}} \otimes {\{14\}} \otimes {\{14\}}= {\{1\}} + \dots$, For the adjoint representation ${\{14\}}$, a color singlet is obtained through combination with one gluon: ${\{14\}} \otimes {\{14\}}= {\{1\}} + \dots$
 For the ${\{27\}}$ representation, combining with two gluons yields the color singlet: ${\{27\}} \otimes {\{14\}} \otimes {\{14\}}= {\{1\}} + \dots$. The gluelump masses for the {\{7\}} and {\{14\}} representations are reported from the lattice correlation functions and are given by \(m_{\{7\}} = 0.46(4)\) and \(m_{\{14\}} = 0.767(5)\). For the \(\{27\}\) representation, using the Casimir scaling ratios from Table~\ref{tab:1}, the gluelump mass can be estimated by scaling the measured mass of the \(\{7\}\) representation:
 
\[
m_{\{27\}} \approx m_{\{7\}} \times \frac{C_{\{27\}}}{C_{\{7\}}} = 0.46 \times \frac{7}{3} \approx 1.07 .
\]

A good approximation for the string breaking energy is given by \(V_{\{r\}}(R) \approx 2m_{{\{r\}}}\). 
Using the above gluelump masses, the corresponding string breaking energies, which determine the asymptotic plateau of the screened potential, are:
\[
V_{\{7\}}(R) \approx 0.92, \qquad
V_{\{14\}}(R) \approx 1.53, \qquad
V_{\{27\}}(R) \approx 2.14
\]
in lattice units. These plateau values are ordered as $V_{\{7\}}(R) < V_{\{14\}}(R) < V_{\{27\}}(R)$, consistent with the increasing 
color charge and string tension of the representations.

Within the model, as shown in Fig.~\ref{fig:2}, the large-distance values of the screened potentials induced by the $z_0$ vacuum domains are consistent with the expected asymptotic behavior. In contrast, as shown in Fig.~\ref{fig:1}, the large-distance values of the potentials induced solely by the $z_1$ vacuum domains do not fully reproduce this ordering of the static potentials, due to unphysical concavity in the intermediate regime that affects the long-distance behavior.

Thus, a model with only \(z_0\) domains exhibits no concavity problems, shows favorable Casimir scaling behavior, and correctly reproduces the expected ordering of the asymptotic potential values with respect to the representation. By contrast, a model with only \(z_1\) domains does not capture these properties fully and deviates from the expected behavior.

As noted by Faber $\it{et~ al.}$ \cite{Faber:1997rp}, it is possible that only domains with the smallest magnitude of center flux have substantial probability. Therefore, the $z_0$ domain model gives a physically-sensible potential.

In the next section, we analyze a possible role of subgroups of the \(G_2\) group in the static potentials obtained by $z_0$ vacuum domains at intermediate distances.

\section{Subgroups of \(G_2\) and the static potentials}\label{Sect3}

The \(G_2\) domain model consisting only of vacuum domains with zero aggregate flux (\(z_0\) domains), which possess the smallest magnitude of center flux, yields the expected static potentials. We now discuss in detail the linear rise of the static potentials at intermediate distances in \(G_2\), which appears in the absence of center vortices. It is therefore of considerable interest to understand the intermediate-range confinement of \(G_2\) gauge theory within the center-vortex picture.

According to lattice simulations \cite{Greensite:2006sm}, removing the gauge links associated with the \(SU(3)\) and \(SU(2)\) subgroups of \(G_2\) leads to either significant deviations from the full \(G_2\) confinement behavior or a complete disappearance of its confinement signatures. These findings indicate that these subgroups play an active role in generating the confining potential. Furthermore, in \(G_2\) gauge-Higgs theory \cite{Holland2003}, introducing a Higgs field in the ${\{7\}}$ representation induces spontaneous symmetry breaking \(G_2\) to \(SU(3)\). This mechanism gives mass to six of the fourteen gluons, while the remaining eight gluons, associated with the unbroken \(SU(3)\) subgroup, stay massless. The resulting theory establishes a smooth interpolation between pure \(G_2\) dynamics and those of its \(SU(3)\) subgroup, revealing how confinement properties evolve under controlled symmetry breaking. The persistent masslessness of the \(SU(3)\) gluons confirms their crucial role as the infrared degrees of freedom responsible for maintaining confinement.

We now investigate a possible role of subgroups in the \(G_2\) potentials of the model. For the \(SU(3)\) and \(SU(2)\) subgroups of \(G_2\) defined in Eqs.~(\ref{H_123}) and (\ref{H_456}), the corresponding center elements are constructed from the centers $Z(3)$ and $Z(2)$, respectively. The fundamental representation $\{7\}$ decomposes under the subgroups as:
\[
\{7\} = \{3\} \oplus \bar{\{3\}} \oplus \{1\} \quad \text{(SU(3))}, \qquad \{7\} = \{2\} \oplus \{2\} \oplus 3\{1\} \quad \text{(SU(2))}.
\]
Using these decompositions, the explicit matrix forms of the center elements for the \(SU(3)\) and \(SU(2)\) subgroups are:
\begin{equation}
\label{Z}
z_{\mathrm{SU(3)}} = 
\begin{pmatrix}
e^{\pm 2\pi i/3} \, \mathbb{I}_{3} & 0 & 0 \\
0 & 1 & 0 \\
0 & 0 & e^{\mp 2\pi i/3} \, \mathbb{I}_{3}
\end{pmatrix},
\qquad
z_{\mathrm{SU(2)}} = 
\begin{pmatrix}
e^{\pi i} \, \mathbb{I}_{2} & 0 & 0 \\
0 & e^{\pi i} \, \mathbb{I}_{2} & 0 \\
0 & 0 & \mathbb{I}_{3}
\end{pmatrix},
\end{equation}
where $\mathbb{I}_{n}$ denotes the $n \times n$ identity matrix.  

We now apply Eq. (\ref{pot:z_0}), substituting the $z_0$ vacuum domains with $z_{SU(3)}$ or $z_{SU(2)}$ center vortices, and employ the fourth \(SU(2)\) subgroup from Eq.~(\ref{H_456}). The real parts of the group factors defined in Eq. (\ref{group_factor}) for the \(SU(3)\) subgroup are identical to those of the \(G_2\) group, as presented in the Appendix. For the \(SU(2)\) subgroup, the real parts of the group factors are also identical to those of the \(G_2\) group but with $\alpha_3^{n} = 0$.

For the $z_{SU(3)}$ center vortices, the angles for all representations are:

\begin{equation}
         \left( {\alpha}_3^{SU(3)}(x) \right)^2 = \frac{A_d}{ 2\mu} \left[
\frac{A}{ A_d} - \frac{A^2}{ A_d^2} \right],~~~
 \\
          \left( {\alpha}_8^{SU(3)}(x)\right)^2 = \frac{A_d}{ 2\mu} \left[
\frac{A}{ A_d} - \frac{A^2}{ A_d^2} \right]
                         + \left(\frac{2\pi\sqrt{24}}{3}\frac{A}{ A_d}\right)^2,                       
\end{equation}

and for the $z_{\mathrm{SU(2)}}$ center vortices, the angle for all representations is:

\begin{equation}
    \left( \alpha_8^{\mathrm{SU(2)}}(x) \right)^2 = \frac{A_d}{ 2\mu} \left[ \frac{A}{ A_d} - \frac{A^2}{ A_d^2} \right] + \left( \pi\sqrt{24} \frac{A}{ A_d} \right)^2.
\end{equation}

\begin{figure}[h!]
\centering
a)\includegraphics[width=0.43\columnwidth]{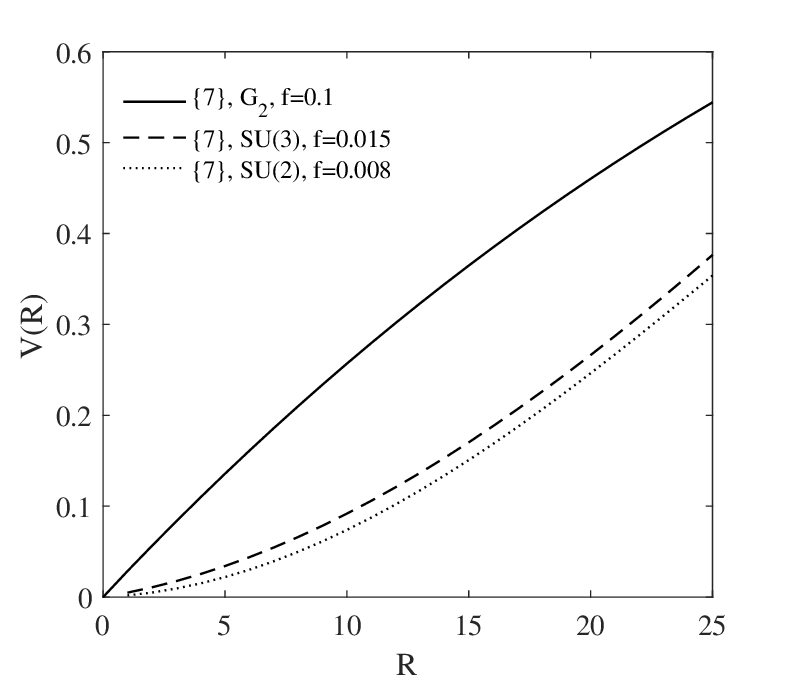}
\hspace{3em}
b)\includegraphics[width=0.43\columnwidth]{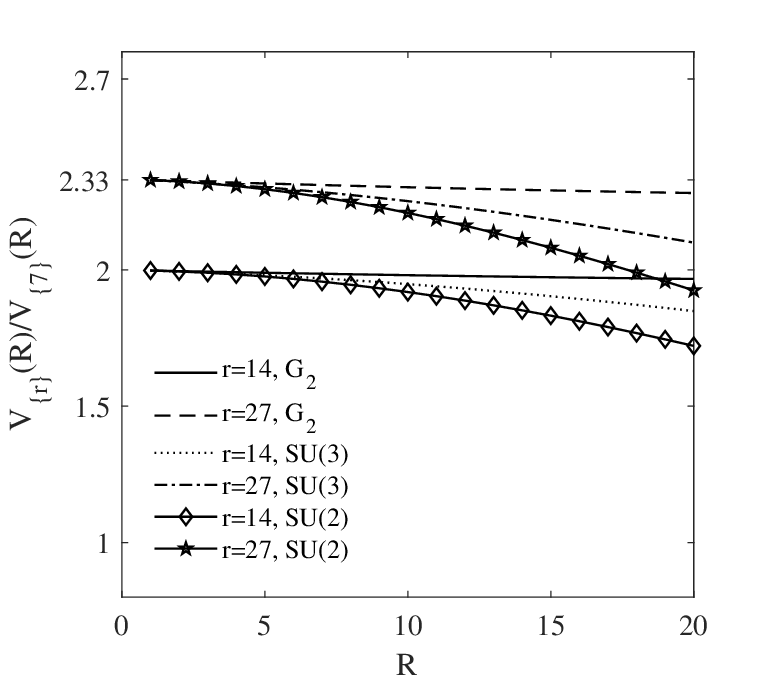}
\caption{a) The static potentials of the ${\{7\}}$ representation for \(G_2\), \(SU(3)\), and \(SU(2)\), fitted to be nearly parallel at intermediate distances with probabilities \(f = 0.1\), \(0.015\), and \(0.008\), respectively. The other fixed parameters are \(L_d = 100\) and \(L_d^2/(2\mu) = 4\). With these fixed probabilities, the potentials for the higher representations, ${\{14\}}$ and ${\{27\}}$, are calculated for each subgroup and are found to be nearly parallel to those of \(G_2\). b) The ratios \(V_{\{r\}}(R)/V_{\{7\}}(R)\) induced by center vortices of each subgroup at intermediate distances in the range \(R \in [1,20]\), compared with those of \(G_2\). The subgroup ratios initially agree with the Casimir ratios of \(G_2\) and exhibit only a slow variation over the range, remaining close to the \(G_2\) values.}\label{fig:G_2-subgroups}
\end{figure}

To investigate the origin of the linear potentials in \(G_2\), the parameters of the model are fixed such that the probability \(f\) that a given plaquette is pierced by a domain is determined by fitting the potentials of the ${\{7\}}$ representation of \(G_2\), \(SU(3)\), and \(SU(2)\) at intermediate distances, yielding values of approximately \(0.1\), \(0.015\), and \(0.008\), respectively, as shown in Fig.~\ref{fig:G_2-subgroups}(a). With these fixed probabilities, the potentials for the higher representations, ${\{14\}}$ and ${\{27\}}$, are calculated for each subgroup and are found to be nearly parallel to those of \(G_2\). Figure~\ref{fig:G_2-subgroups}(b) displays the potential ratios \(V_{\{14\}}(R)/V_{\{7\}}(R)\) and \(V_{\{27\}}(R)/V_{\{7\}}(R)\) induced by center vortices of the subgroups, compared with those of \(G_2\) induced by \(z_0\) vacuum domains. These subgroup ratios initially match the Casimir ratios provided in Table~\ref{tab:1} and vary only slowly over the range \(R \in [1,20]\), remaining close to the values obtained for \(G_2\).

Consequently, the potential ratios of the \(SU(3)\) and \(SU(2)\) subgroups, which remain close to those of \(G_2\), may indicate their possible role in the linear rise of the static potentials at intermediate distances in \(G_2\).

\section{Conclusion}\label{Sect4}

Within the framework of the domain model of confinement, we have investigated the domain structures of \(G_2\) gauge theory and the associated Yang-Mills static potential. Since the \(G_2\) group has a trivial center, vacuum domains with non-zero net magnetic flux (\(z_1\) domains) as well as domains with zero net magnetic flux (\(z_0\) domains) may exist. The thickness of vacuum domains is therefore important for understanding the intermediate-range confinement regime.

We find that a model featuring only \(z_1\) domains violates the convexity of the static potential and does not exhibit favorable Casimir scaling properties. Moreover, while all representations are screened at asymptotic distances due to the absence of a non-trivial center in \(G_2\), the \(z_1\) model deviates from the expected ordering of the asymptotic potential values with respect to the representation. This deviation is attributed to unphysical concavity in the intermediate regime, which affects the long-distance behavior.

In contrast, a model consisting exclusively of \(z_0\) domains exhibits no concavity problems, displays favorable Casimir scaling, and correctly reproduces the expected ordering of the asymptotic potential values with respect to the representation, consistent with the plateau values determined by the corresponding gluelump masses. It is plausible that only domains with the smallest magnitude of center flux carry substantial probability. The \(z_0\) domain model thus yields a physically sensible potential.

We have also analyzed the possible role of subgroups of the \(G_2\) group in the linear rise of the static potentials of \(G_2\) at intermediate distances. To investigate this, the parameters of the domain model are fixed by fitting the \(\{7\}\)-representation potentials of \(G_2\), \(SU(3)\), and \(SU(2)\) at intermediate distances. With these fixed probabilities, the potentials for the higher representations, \(\{14\}\) and \(\{27\}\), are calculated for each subgroup and are found to be nearly parallel to those of \(G_2\). The potential ratios of the \(SU(3)\) and \(SU(2)\) subgroups remain close to those of \(G_2\), suggesting that these subgroups may play a role in the linear rise of the static potentials at intermediate distances in \(G_2\).

\appendix
\section{Cartan Generators and Group Factors for \(G_2\) Representations}

The Cartan generators $H_8^{\{r\}}$ and $H_3^{\{r\}}$ for representation $r$ appearing in the group factors of the static potential given in Eq.~(\ref{potential}) can be calculated using the decompositions of the representations. The real part of the group factors can be obtained using both Cartan generators $H_8^{\{r\}}$ and $H_3^{\{r\}}$ for the $z_n$ ($n=0,1$) vacuum domains in several representations as follows:

\subsection{Fundamental ($\{7\}$) Representation}
The fundamental representation of \(G_2\) decomposes under its \(SU(3)\) subgroup as $\{7\} = \{3\} \oplus \bar{\{3\}} \oplus \{1\}$. The Cartan generators in this basis are:
\begin{equation}
\begin{aligned}
\label{eq:H7}
H_3^{\{7\}} &= \frac{1}{2\sqrt{2}} \big( P_{1,1} - P_{2,2} - P_{5,5} + P_{6,6} \big), \\
H_8^{\{7\}} &= \frac{1}{2\sqrt{6}} \big( P_{1,1} + P_{2,2} - 2P_{3,3} - P_{5,5} - P_{6,6} + 2P_{7,7} \big),
\end{aligned}
\end{equation}
where $(P_{i,j})_{\alpha \beta} = \delta_{i \alpha} \delta_{j \beta}$, and $\alpha, \beta$ indicate the row and column of the matrices, respectively. The real part of the group factor in representation $\{7\}$ is:
\begin{equation}
\mathrm{Re}\,G_{\{7\}}(\alpha^{n}) = \frac{1}{7}\left[2\cos\left(\frac{\alpha_3^{n}}{2\sqrt{2}} + \frac{\alpha_8^{n}}{2\sqrt{6}}\right) + 2\cos\left(-\frac{\alpha_3^{n}}{2\sqrt{2}} + \frac{\alpha_8^{n}}{2\sqrt{6}}\right) + 2\cos\left(\frac{\alpha_8^{n}}{\sqrt{6}}\right) + 1\right].
\label{group-7}                      
\end{equation}

\subsection{Adjoint ($\{14\}$) Representation}
The adjoint representation of \(G_2\) decomposes under its \(SU(3)\) subgroup as $\{14\} = \{8\} \oplus \{3\} \oplus {\{\bar{3}\}}$. The Cartan generators in this basis are:
\begin{equation}
\begin{aligned}
\label{eq:H14}
H_3^{\{14\}} &= \frac{1}{2\sqrt{2}} \big( P_{1,1} - P_{2,2} - P_{4,4} + P_{5,5} + 2P_{8,8} + P_{9,9} - 2P_{10,10} - P_{12,12} - P_{13,13} + P_{14,14} \big), \\
H_8^{\{14\}} &= \frac{1}{2\sqrt{6}} \big( P_{1,1} + P_{2,2} - 2P_{3,3} - P_{4,4} - P_{5,5} + 2P_{6,6} + 3P_{9,9} + 3P_{12,12} - 3P_{13,13} - 3P_{14,14}\big),
\end{aligned}
\end{equation}
and the real part of the group factor in representation $\{14\}$ is:
\begin{equation}
\begin{aligned}
\mathrm{Re}\,G_{\{14\}}(\alpha^{n}) &= \frac{1}{14}\bigg[2\cos\left(\frac{\alpha_3^{n}}{2\sqrt{2}} + \frac{\alpha_8^{n}}{2\sqrt{6}}\right) + 2\cos\left(-\frac{\alpha_3^{n}}{2\sqrt{2}} + \frac{\alpha_8^{n}}{2\sqrt{6}}\right) \\
&\quad + 2\cos\left(\frac{\alpha_8^{n}}{\sqrt{6}}\right) + 2\cos\left(\frac{\alpha_3^{n}}{\sqrt{2}}\right) + 2\cos\left(\frac{\alpha_3^{n}}{2\sqrt{2}} + \frac{3\alpha_8^{n}}{2\sqrt{6}}\right) \\
&\quad + 2\cos\left(\frac{\alpha_3^{n}}{2\sqrt{2}} - \frac{3\alpha_8^{n}}{2\sqrt{6}}\right) + 2\bigg].
\label{group-14} 
\end{aligned}                     
\end{equation}

\subsection{$\{27\}$ Representation}
The $\{27\}$ representation of \(G_2\) decomposes under its \(SU(3)\) subgroup as $\{27\} = {\{8\}} \oplus {\{6\}} \oplus {\{\bar{6}\}} 
\oplus {\{3\}} \oplus {\{\bar{3}\}} \oplus {\{1\}}$. The Cartan generators in this basis are:
\begin{equation}
\begin{aligned}
\label{eq:H27}
H_3^{\{27\}} &= \frac{1}{2\sqrt{2}} \big( P_{1,1} - P_{2,2} - P_{4,4} + P_{5,5} + 2P_{8,8} + P_{9,9} - 2P_{10,10} - P_{12,12} - P_{13,13} + P_{14,14} \\
&\quad + 2P_{15,15} + P_{17,17} - 2P_{18,18} - P_{19,19} - 2P_{21,21} - P_{23,23} + 2P_{24,24} + P_{25,25} \big), \\
H_8^{\{27\}} &= \frac{1}{2\sqrt{6}} \big( P_{1,1} + P_{2,2} - 2P_{3,3} - P_{4,4} - P_{5,5} + 2P_{6,6} + 3P_{9,9} + 3P_{12,12} - 3P_{13,13} - 3P_{14,14} \\
&\quad + 2P_{15,15} + 2P_{16,16} - P_{17,17} + P_{18,18} - P_{19,19} - 4P_{20,20} - 2P_{21,21} - 2P_{22,22} + P_{23,23} \\ 
&\quad - P_{24,24} + P_{25,25} + 4P_{26,26}\big),
\end{aligned}                     
\end{equation}
and the real part of the group factor in representation $\{27\}$ is:
\begin{equation}
\begin{aligned}
\mathrm{Re}\,G_{\{27\}}(\alpha^{n}) &= \frac{1}{27} \bigg[ 2\cos\left(\frac{\alpha_3^{n}}{2\sqrt{2}} + \frac{\alpha_8^{n}}{2\sqrt{6}}\right) + 2\cos\left(-\frac{\alpha_3^{n}}{2\sqrt{2}} + \frac{\alpha_8^{n}}{2\sqrt{6}}\right) \\
&\quad + 2\cos\left(\frac{\alpha_8^{n}}{\sqrt{6}}\right) + 2\cos\left(\frac{\alpha_3^{n}}{\sqrt{2}}\right) + 2\cos\left(\frac{\alpha_3^{n}}{2\sqrt{2}} + \frac{3\alpha_8^{n}}{2\sqrt{6}}\right) \\
&\quad + 2\cos\left(\frac{\alpha_3^{n}}{2\sqrt{2}} - \frac{3\alpha_8^{n}}{2\sqrt{6}}\right) + 2\cos\left(\frac{\alpha_3^{n}}{\sqrt{2}} + \frac{\alpha_8^{n}}{\sqrt{6}}\right) \\
&\quad + 2\cos\left(\frac{\alpha_8^{n}}{\sqrt{6}}\right) + 2\cos\left(\frac{\alpha_3^{n}}{2\sqrt{2}} - \frac{\alpha_8^{n}}{2\sqrt{6}}\right) + 2\cos\left(-\frac{\alpha_3^{n}}{\sqrt{2}}+ \frac{\alpha_8^{n}}{\sqrt{6}}\right)\\
&\quad + 2\cos\left(\frac{\alpha_3^{n}}{2\sqrt{2}} + \frac{\alpha_8^{n}}{2\sqrt{6}}\right)+ 2\cos\left(\frac{2\alpha_8^{n}}{\sqrt{6}}\right) + 3 \bigg].
\label{group-27} 
\end{aligned}                     
\end{equation}

\end{document}